\documentclass[5p,times,twocolumn,left=1in, right=1in]{elsarticle}
\journal{Physica D}

\usepackage[utf8]{inputenc}
\usepackage{amsfonts} 
\usepackage{amsmath}
\usepackage{amssymb}
\usepackage[hyperref]{xcolor}
\usepackage{bm}
\usepackage[hyphens]{xurl}
\usepackage{hyperref}
\usepackage{algorithm}
\usepackage{algpseudocode}
\usepackage{booktabs}
\renewcommand{\v}[1]{\boldsymbol{#1}}
\renewcommand{\d}{\,\mathrm{d}}

\newcommand{\added}[1]{{\color{black}#1}}
    \hypersetup{colorlinks=true}

\begin{document}
\hypersetup{
  linkcolor=blue,
  urlcolor=blue,
  citecolor=blue
}
\title{KGMM: A K-means Clustering Approach to Gaussian Mixture Modeling for Score Function Estimation}




\author[LGT]{Ludovico T Giorgini\corref{cor1}}
\ead{ludogio@mit.edu}         
\ead[url]{ludogiorgi.github.io}
\affiliation[LGT]{
                 organization={Department of Mathematics},
                 addressline={MIT},
                 city={Cambridge},
                 state={MA},
                 country={USA}}
\cortext[cor1]{Corresponding author}

\author[TB]{Tobias Bischoff}
\affiliation[TB]{
                 organization={Aeolus Labs},
                 city={San Francisco},
                 state={CA},
                 country={USA}}

\author[AS]{Andre N Souza}
\ead{sandre@mit.edu}    
\ead[url]{sandreza.github.io}
\affiliation[AS]{
                 organization={Department of Earth, Atmospheric, and Planetary Sciences},
                 addressline={MIT},
                 city={Cambridge},
                 state={MA},
                 country={USA}}

\begin{abstract}
We propose a hybrid method for accurately estimating the score function, i.e., the gradient of the log steady‑state density, using a Gaussian Mixture Model (GMM) in conjunction with a bisecting K-means clustering step. Our approach, which we call KGMM, offers a systematic way to combine statistical density estimation with a neural-network-based interpolation of the score, leveraging the strengths of both. We demonstrate its ability to accurately reconstruct the long-time statistical properties of several paradigmatic systems, including potential systems\added{,} and chaotic Lorenz-type models\added{, and the Kuramoto--Sivashinsky equation}. Numerical experiments show that KGMM yields robust estimates of the score function, even for small values of the covariance amplitude in the GMM, where the standard GMM methods tend to fail because of noise amplification. \added{We compare the performance of KGMM against the conventional Denoising Score Matching (DSM) approach, demonstrating that KGMM achieves more faithful reconstruction of the steady-state distribution for low-dimensional systems at a fraction of the computational cost.} These accurate estimates allow us to build effective stochastic reduced-order models that reproduce the invariant measures of the target dynamics.
\end{abstract}

\maketitle

    \textcolor{red}{
    } 




\section{Introduction}
The score function, defined as the gradient of the logarithm of a system's steady-state probability density function, is a fundamental quantity in statistical physics, dynamical systems, and machine learning. It underpins key theoretical frameworks such as the Generalized Fluctuation-Dissipation Theorem (GFDT) \cite{marconi2008,giorgini_response_theory, giorgini2025predicting,cooper2011climate,baldovin2020understanding,ghil2020physics}, which links spontaneous fluctuations to system responses, and plays a crucial role in generative modeling \cite{song2021}, parameter estimation \cite{silverman1986}, and causal inference \cite{shimizu2007}. Crucially, knowledge of the score function provides insights into the dynamical features of a system without requiring explicit knowledge of its governing equations. Instead, it can be inferred from statistical properties, which are often more accessible in experimental and numerical settings \citep{falasca2024data, giorgini2025learning, giorgini2024reduced, souza2024representing_a, souza2024representing_b}.

Accurate and efficient estimation of the score function remains a formidable challenge, particularly in high-dimensional systems. Gaussian Mixture Models (GMMs) are widely used to approximate complex probability distributions due to their flexibility and well-established probabilistic framework \cite{teh2006}. In a GMM, the probability density function is modeled as a weighted sum of Gaussian components, where the mean vectors of the Gaussians are chosen to span the state space explored by the underlying dynamical system. 

A critical aspect of using GMMs is the selection of the covariance matrix amplitude for each Gaussian component. Larger covariance amplitudes result in a smoother estimated invariant density because the Gaussian kernel effectively averages out local fluctuations. However, this smoothing comes at a cost: the estimated density is perturbed relative to the true invariant density, as the convolution with the Gaussian kernel tends to blur finer details of the distribution. Conversely, smaller covariance amplitudes produce an invariant density estimate that more closely resembles the true distribution. Yet, the reduction in smoothing increases the noise level in the estimate, which is particularly problematic when differentiating the density to compute the score function. Here, even slight noise amplification can lead to significant inaccuracies in the gradient estimates. Although increasing the number of Gaussian mixture components can help mitigate these issues by providing a more detailed approximation, this solution introduces additional computational burdens and an elevated risk of overfitting \cite{reynolds2009}.

Recent advancements in score-based generative modeling \cite{hyvarinen2005estimation, vincent2011connection, song2021, vargas2023bayesian, bischoff2024enhancing, schwank2025robust} offer an alternative strategy by directly training a neural network to approximate the score function via a dataset-wide loss minimization procedure, commonly known as Denoising Score Matching (DSM). This method relies on the implicit regularization afforded by the neural network training procedure to define a ``smoothed" version of the Gaussian mixture score function. However, this approach is computationally expensive, as the loss function depends on the entire dataset, and there is no guarantee that the learned score function converges to the true underlying gradient field.

In this work, we propose a hybrid approach that leverages both GMM-based statistical estimation and neural network interpolation. Our method first computes the score function at representative points in the state space by combining a bisecting K-means clustering algorithm with GMM. As we will show, this strategy enables the efficient evaluation of a discretized version of the score function by leveraging information from the entire dataset. We then train a neural network to interpolate between these points. This method combines the advantages of probabilistic density modeling with the flexibility of machine learning, leading to a computationally efficient and precise framework for score function estimation in large datasets.

The article is structured as follows. Section 2 presents the KGMM method, detailing its derivation and advantages over standard GMMs. Section 3 validates KGMM through numerical experiments on potential and chaotic systems, comparing estimated score functions with analytical solutions when available\added{, and demonstrates scalability on the Kuramoto--Sivashinsky equation in dimensions up to 16}. \added{Section 4 compares the computational performance of KGMM-preprocessed training versus direct neural network training and discusses the method's limitations and practical guidelines for hyperparameter selection.} Section \added{5} concludes with key findings and future directions.

\section{Method}

\subsection{Motivation}
The dynamics of physical systems often exhibit a hierarchical structure in their spatiotemporal evolution, wherein predictable, low-dimensional processes emerge on longer timescales and larger spatial scales, while chaotic, high-dimensional fluctuations dominate at shorter timescales and finer spatial resolutions. In many complex systems, the details of small-scale, fast processes become increasingly irrelevant under coarse-graining transformations and can be effectively replaced by stochastic forcing terms that preserve essential statistical and dynamical properties. This paradigm not only provides a faithful representation of the underlying physics but also enables a significant reduction in the dimensionality of high-dimensional systems, facilitating both analytical tractability and numerical efficiency.

A paradigmatic example of this approach is found in climate physics, where large-scale, slow dynamics, such as ocean circulation and seasonal variations, coexist with small-scale, rapid processes, including turbulent eddies and convective storms. Reduced-order stochastic models provide an effective means of capturing the statistical and dynamical structure of such multiscale interactions, successfully replicating phenomena like the El Niño-Southern Oscillation (ENSO), monsoonal cycles, and long-range teleconnections, as well as the coupling of climate variables observed in paleoclimate data \citep{majda_applied_math, chen2023rigorous, keyes2023stochastic, giorgini2022non, baldovin2022extracting}.

Based on observations of a physical system characterized by a steady-state distribution $\rho_S(\textbf{x})$ and a time correlation function $\bm{C}(t)$, the following Langevin equation is constructed to inherently reproduce these properties:
\begin{equation}
    \dot{\bm{x}}(t) = \bm{\Sigma} \bm{\Sigma}^T \nabla \ln \rho_S(\bm{x}) + \sqrt{2}\bm{\Sigma} \bm{\xi}(t),
    \label{eq:langevin}
\end{equation}
where $\bm{\xi}(t)$ is a vector of independent Gaussian white noise processes, and the covariance matrix $\bm{\Sigma}$ is chosen to match the time-correlations of the observed data. This formulation ensures that $\rho_S$ remains invariant under the corresponding Fokker-Planck operator,
\begin{equation}
    \mathcal{L}_{FP} \rho_S = 0, \quad \text{with} \quad \mathcal{L}_{FP}f = - \nabla \cdot \left( \bm{\Sigma} \bm{\Sigma}^T \nabla \ln \rho_S f \right) + \nabla \cdot \left( \bm{\Sigma} \bm{\Sigma}^T \nabla f  \right),
\end{equation}
which governs the evolution of the probability density in the reduced-order model.

The key observation here is that the deterministic drift term in the Langevin equation (\ref{eq:langevin}) is determined by the score function, $\nabla \ln \rho_S(\bm{x})$, which encapsulates the structure of the underlying dynamical system. Knowledge of this drift term provides insight into the statistical and dynamical properties of the observed system, including the ability to quantify how the system responds to external perturbations \cite{baldovin2021handy}. In the next section, we will show how, by leveraging statistical estimation techniques alongside machine learning approaches, it becomes possible to reconstruct this fundamental quantity with high fidelity, offering new avenues for the systematic derivation of stochastic models in complex dynamical systems.

\subsection{Derivation of the Score Function}
A Gaussian Mixture Model (GMM) models a probability density function as a weighted sum of Gaussian components:
\begin{equation}
    p_\sigma(\bm{x}) = \sum_{k=1}^{K} w_k \mathcal{N}(\bm{x} \mid \bm{\mu}_k, \bm{\Sigma}_k),
\end{equation}
where $\rho_S(\bm{x})$ denotes the stationary density of the underlying dynamical system, $w_k$ denotes the weights representing the probability associated with each component $k$, $\bm{\mu}_k$ denotes the mean vectors, and the covariance matrices are assumed to be isotropic, i.e., $\bm{\Sigma}_k = \sigma^2 \bm{I}$, with $\bm{I}$ as the identity matrix. The weights $w_k$ indicate the proportion of the dataset that each $\bm{\mu}_k$ represents; they sum to one.

The score function, defined as the gradient of the logarithm of the probability density, is given by:
\begin{equation}
    \nabla \ln p_\sigma(\bm{x}) = -\frac{1}{\sigma^2} \sum_{k=1}^{K} \frac{w_k \mathcal{N}(\bm{x} \mid \bm{\mu}_k, \sigma^2 \bm{I}) (\bm{x} - \bm{\mu}_k)}{p_\sigma(\bm{x})}.
    \label{eq:score_disc}
\end{equation}
\added{We now specialize the expression for the score to the case where $K = N$, corresponding to the number of data points. This choice is central to our method and should not be confused with $N_C$, the number of clusters used for aggregation, which we introduce later and satisfies $N_C \ll N$.}
Defining the change of variables
\begin{equation}
    \bm{z}_k = \bm{x} - \bm{\mu}_k,
\end{equation}
and taking the limit $N\to \infty$, we can \added{formally} rewrite Eq. \eqref{eq:score_disc} as
\begin{equation}
    \nabla \ln p_\sigma(\bm{x}) \added{\approx} -\frac{1}{\sigma^2} \int_{\Omega_{\bm{\mu}}} \frac{p(\bm{\mu})\mathcal{N}(\bm{z} \mid \bm{0}, \sigma^2 \bm{I})}{p_\sigma(\bm{x})}\bm{z}\d \bm{\mu},
    \label{eq:score_cont}
\end{equation}
where the integral is carried out over the whole phase space $\Omega_{\bm{\mu}}$ and \added{we approximate} $p(\bm{\mu})\added{\approx}\rho_S(\bm{\mu})$\added{, i.e., we assume the empirical distribution of data points approximates the true invariant measure. This is the first approximation in our method and is valid in the regime of large $N$ when the data points are sampled from $\rho_S$.}
Let us \added{now} define
\begin{equation}
    p(\bm{z}) = \mathcal{N}(\bm{z} \mid \bm{0}, \sigma^2 \bm{I})
\end{equation}
\added{as} the probability density function of $\bm{z}$\added{, and} rewrite the probability density function of $\bm{\mu}$ as
\begin{equation}
    p(\bm{\mu}) = p(\bm{\mu}+\bm{z}\mid\bm{z})=p(\bm{x}\mid\bm{z}).
\end{equation}
Thus, we can express
\begin{equation}
     \frac{p(\bm{x} \mid \bm{z}) p(\bm{z})}{p_\sigma(\bm{x})} = p(\bm{z} \mid \bm{x}),
\end{equation}
\added{since $p_\sigma(\bm{x})$ is the marginal of $\bm{x}$ under the Gaussian perturbation model. }
Substituting this back into the score expression, we obtain
\begin{equation}
    \nabla \ln p_\sigma(\bm{x}) \added{\approx} -\frac{1}{\sigma^2} \int_{\Omega_{\bm{\mu}}} p(\bm{z} \mid \bm{x}) \bm{z} \d \bm{z} = -\frac{1}{\sigma^2}\mathbb{E}[\bm{z}\mid\bm{x}].
\end{equation}

\added{The consistency of this approximation can be understood in two limiting regimes: (i) As $N \to \infty$ with fixed $\sigma$, the empirical distribution of $\{\bm{\mu}_i\}$ converges to $\rho_S$, and $p_\sigma$ becomes a well-defined convolution of $\rho_S$ with a Gaussian kernel. (ii) As $\sigma \to 0$ with fixed $N$, the Gaussian kernels become increasingly localized, and $p_\sigma \to \rho_S$ pointwise where data is available. In practice, we work with finite $N$ and finite $\sigma$, introducing a controlled bias that is regularized by the subsequent neural network interpolation.}

We evaluate the score function at a finite set of points in phase space. To this end, we partition the phase space into $N_C$ clusters $\{\Omega_j\}_{j=1}^{N_C}$ with corresponding centroids $\bm{C}_j$.

The number of clusters, \(N_C\), introduces a critical performance trade-off. A larger \(N_C\) improves the spatial resolution of score function estimates by allowing finer-grained cluster subdivisions that better approximate the local gradient structure near the centroids. However, an excessively large \(N_C\) reduces the number of samples per cluster, which amplifies statistical noise in the averaged score estimates, while too few clusters risk oversmoothing the score function—particularly in regions of rapid gradient variation. Moreover, in high-dimensional spaces, the exponential growth of the feature space necessitates a careful increase in \(N_C\) with the dimension \(d\); finer subdivisions become essential to capture local variations without loss of information. Empirically, one may adopt a scaling rule of the form
\begin{equation}
N_C \propto \sigma^{-d},
\end{equation}
where \(\sigma\) denotes the covariance amplitude. This scaling ensures that each cluster is sufficiently homogeneous for accurate estimation while still containing enough data points, thereby balancing spatial resolution with statistical reliability. Optimal \(N_C\) is ultimately guided by both the characteristic length scales of the underlying density, \(\rho_S(\mathbf{x})\), and the intrinsic dimensionality of the dataset.

We use the bisecting K-means clustering algorithm of \cite{souza2024modified}. The bisecting K-means algorithm was selected over density-based methods such as DBSCAN~\cite{ester1996density} due to its deterministic partitioning behavior and scalability in high-dimensional spaces. While DBSCAN excels at identifying arbitrarily shaped clusters with minimal parameter tuning, its reliance on neighborhood density calculations becomes computationally prohibitive for large $N$-dimensional datasets. In contrast, bisecting K-means achieves a time complexity of $\mathcal{O}(N \cdot D \cdot \log N_C)$ in $D$ dimensions through iterative binary splits, thus avoiding the pairwise distance comparisons of $\mathcal{O}(N^2)$ that are inherent to density-based approaches. This hierarchical strategy effectively preserves cluster coherence in sparse regions while maintaining linear scalability with dataset size—an essential advantage when processing large samples.

The average score within each cluster is then given by
\begin{equation}
    \nabla \ln p(\bm{C}_j) \approx -\frac{1}{\sigma^2}\int_{\Omega_j} \mathbb{E}[\bm{z}\mid\bm{x}] p(\bm{x}) d\bm{x}.
\end{equation}
This integral is approximated by summing over sample values of $\bm{x}$ drawn from $p(\bm{x})$ within each cluster, and normalizing by the number of samples in the cluster, denoted $N_C^j$. In our implementation, we generate these sample points by drawing $N$ samples using
\begin{equation}
    \bm{x}_i = \bm{\mu}_i + \bm{z}_i,
    \label{eq:x_mu_z}
\end{equation}
where $\bm{\mu}_i$ are the data points and $\bm{z}_i$ are random variables drawn from $\mathcal{N}(0,\sigma^2 \bm{I})$. Thus, the discretized form of the K-means cluster-averaged GMM score function (KGMM) becomes
\begin{equation}
    \nabla \ln p(\bm{C}_j) \approx -\frac{1}{N_C^j\sigma^2} \sum_{i: \bm{x}_i \in \Omega_j} \bm{z}_i = \frac{\bm{q}_j}{\sigma}.
\end{equation}

This procedure can be iterated by repeatedly generating new samples $\bm{x}_i$ using the same data points $\bm{\mu}_i$ along with newly drawn noise vectors $\bm{z}_i$. Subsequently, a neural network is employed to interpolate between the computed cluster-wise estimates $\nabla \ln p(\bm{C}_j)$, yielding a continuous approximation of the score function. The neural network $\bm{q}_{\theta}$ is trained to minimize the following loss function:
\begin{equation}
    \mathcal{L}(\theta) = \frac{1}{N_C} \sum_{k=1}^{N_C} \left\| \bm{q}_{\theta}(\bm{C}_k) - \bm{q}_k \right\|_2^2,
    \label{loss_kmeans}
\end{equation}
where $\bm{q}_k$ is our cluster-wise estimate of $-\mathbb{E}[\bm{z}|\bm{x}]$ with $\bm{x}, \bm{z}$ defined in Eq. \eqref{eq:x_mu_z}.

The complete procedure is summarized in Algorithm~\ref{alg:score_estimation}.

\begin{algorithm}
\caption{KGMM Score Function Estimation}
\label{alg:score_estimation}
\begin{algorithmic}[1]
\Require Dataset $\{\bm{\mu}_i\}_{i=1}^{N}$, number of clusters $N_C$, noise level $\sigma$, convergence threshold $\alpha$
\State \added{// Note: In the GMM formulation, $K = N$ mixture components, but here we aggregate into $N_C \ll N$ clusters}
\State Initialize K-means clustering to partition $\{\bm{\mu}_i\}$ into $\{\Omega_k\}_{k=1}^{N_C}$ with centroids $\{\bm{C}_k\}$
\Repeat
    \For{$i = 1$ to $N$}
        \State Generate noise $\bm{z}_i \sim \mathcal{N}(\bm{0}, \sigma^2\bm{I})$
        \State Compute perturbed point $\bm{x}_i = \bm{\mu}_i + \bm{z}_i$
        \State Assign $\bm{x}_i$ to cluster $\Omega_k$
    \EndFor
    \For{$k = 1$ to $N_C$}
        \State Compute $\bm{q}_k = -\frac{1}{|\Omega_k|\sigma} \sum_{i \in \Omega_k} \bm{z}_i$
    \EndFor
\Until{Convergence criterion $\|\bm{q}_k^{(t)} - \bm{q}_k^{(t-1)}\| < \alpha$ for all $k$}
\State Train neural network parameters $\theta$ by minimizing loss $\mathcal{L}(\theta)$ in Eq.~\eqref{loss_kmeans}
\end{algorithmic}
\label{algorithm}
\end{algorithm}

\added{\subsection{Relation to Denoising Score Matching}}
\added{Our KGMM method shares conceptual similarities with Denoising Score Matching (DSM) \cite{vincent2011connection, hyvarinen2005estimation}, which has become a cornerstone of modern score-based generative models \cite{song2021}. In DSM, one perturbs data with Gaussian noise and trains a neural network to predict the noise vector, effectively learning the score of the noise-perturbed distribution. The key insight is that the score of a Gaussian-convolved density can be estimated more easily than the score of the original density.}

\added{Specifically, DSM considers data $\bm{x}_0 \sim \rho_S$ and perturbed samples $\bm{x} = \bm{x}_0 + \bm{z}$ where $\bm{z} \sim \mathcal{N}(\bm{0}, \sigma^2\bm{I})$. The DSM objective minimizes}
\begin{equation}\added{
    \mathcal{L}_{\text{DSM}}(\theta) = \mathbb{E}_{\bm{x}_0 \sim \rho_S, \bm{z} \sim \mathcal{N}(\bm{0}, \sigma^2 \bm{I})} \left[ \left\| \bm{s}_\theta(\bm{x}_0 + \bm{z}) + \frac{\bm{z}}{\sigma^2} \right\|^2 \right],}
\label{eq:dsm_objective}
\end{equation}
\added{where $\bm{s}_\theta$ is a neural network. This is equivalent to learning $\nabla_{\bm{x}} \log p_\sigma(\bm{x})$, the score of the convolved distribution $p_\sigma(\bm{x}) = \int \rho_S(\bm{x}_0) \mathcal{N}(\bm{x} \mid \bm{x}_0, \sigma^2 \bm{I}) \d\bm{x}_0$.}

\added{KGMM can be viewed as a two-stage approach that first estimates the conditional expectation $\mathbb{E}[\bm{z} \mid \bm{x}] = -\sigma \nabla_{\bm{x}} \log p_\sigma(\bm{x})$ at cluster centers using the GMM construction, and then interpolates these estimates with a neural network. The key distinction is that KGMM leverages explicit statistical estimation via clustering to compute score estimates at representative points before neural network interpolation, whereas DSM directly trains on the full dataset. This distinction leads to computational advantages for large $N$, as we demonstrate in Section 4. Both methods share the finite-$\sigma$ bias: as $\sigma \to 0$, the score of $p_\sigma$ approaches the score of $\rho_S$, but for finite $\sigma$, the convolution introduces smoothing that can blur sharp features of the true density.}

\subsection{Illustrative Example: KGMM vs. GMM in One Dimension}
In this subsection, we compare the score function obtained via the standard GMM approach with the one using the proposed KGMM algorithm, highlighting how KGMM remains accurate even for small covariance amplitudes $\sigma$. To illustrate the differences, we consider the one-dimensional system
\begin{align}
\dot{x}(t) = x - x^3 + \sqrt{2} \xi(t),
\end{align}
with $\xi(t)$ delta-correlated Gaussian white noise. This system has the exact score function $s(x) = x - x^3$ and density $\rho \propto e^{-U}$, where $U(x) = (1-x^2)^2/4$.

We use $N_{\text{eff}} = 10^5$ effectively uncorrelated samples from the distribution $\rho$, denoted by $\mu_\omega$, and fit a Gaussian mixture model of the form
\begin{align}
\rho(x) = \frac{1}{N} \sum_{\omega = 1}^N \frac{1}{\sqrt{2 \pi \sigma^2}} e^{\frac{-(x - \mu_\omega)^2}{2\sigma^2}}.
\end{align}
The corresponding GMM score function for various choices of $\sigma$ is
\begin{align}
\nabla \ln \rho (x) = \frac{\sum_{\omega = 1}^N (\mu_\omega - x) e^{\frac{-(x - \mu_\omega)^2}{2\sigma^2}}}{\sigma^2 \sum_{\omega = 1}^N e^{\frac{-(x - \mu_\omega)^2}{2\sigma^2}}}.
\end{align}

\begin{figure*}
 	\centering	\includegraphics[width=\textwidth]{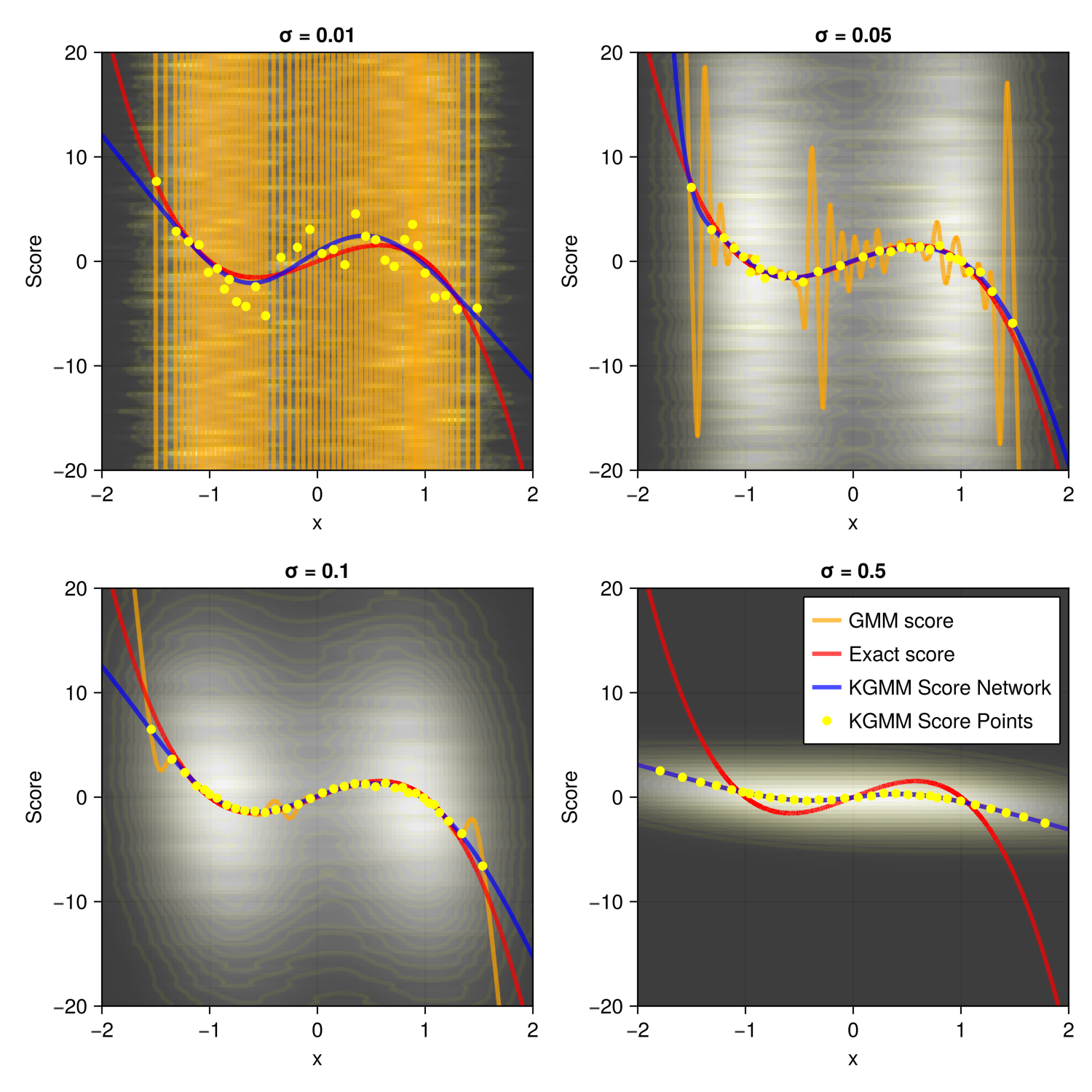}
 	\caption{Comparison for different values of $\sigma$ between the score function obtained through the standard GMM (orange curve) and the one (blue curve) obtained by interpolating the discrete values of the KGMM score function (yellow points). Note that for small $\sigma$, the standard GMM curve becomes significantly noisier, whereas the KGMM approach preserves a close agreement with the true score (red curve). Each panel's white and black background represents the joint distribution of  $(x_\omega, -z_\omega/\sigma),\, \omega \in \{1,\cdots N\}$. Fixing a value of $x$ and computing the expected value of the resulting conditional density yields the value of the yellow points. }
 	\label{fig:score_estimate}
\end{figure*}

To apply KGMM, we then draw $N$ samples of a random normal variable $Z_\omega$, $\omega = 1, ..., N$, and construct
\begin{align}
x_\omega = \mu_\omega + z_\omega.
\end{align}
We formulate the joint density $(x, z)$, cluster each $x_\omega$ into $N_C \approx 30$ clusters via K-means, assign the same cluster of $x_\omega$ to $z_\omega$, average each $z_\omega$ over a cluster, and divide by $-\sigma^2$, ultimately learning a discrete approximation of the score function that is then interpolated by a neural network. This describes only one iteration of Algorithm~\ref{alg:score_estimation} since we perturb each data point with noise only once.  See Figure \ref{fig:score_estimate} for an illustration of this procedure for various choices of $\sigma$.  More generally, we would construct $x_{\omega \omega'} = \mu_\omega + z_{\omega'}$ and iterate both $\omega \in \{1, ..., N \}$ and $\omega' \in \{1, ..., N \times M \}$ for some natural number $M \geq 1$ until we have a converged estimate of the score. 

When the amplitude of the covariance matrix $\sigma$ in the standard GMM is decreased, we obtain a noisier estimation of the score function because the differentiation becomes more sensitive to data fluctuations. By contrast, our KGMM algorithm leverages the additional cluster-based regularization and the subsequent neural network interpolation to remain stable for small values of $\sigma$, achieving good agreement with the true score function. 

\section{Results}
We tested the proposed KGMM score estimation algorithm on \added{five} different stochastic reduced-order models relevant in climate science \added{and chaotic dynamics}. For each system, we constructed the score function using KGMM and compared it with its analytic expression when available. We also used the estimated KGMM score function to generate stochastic trajectories by integrating \added{Eq.~(\ref{eq:langevin}) with $\bm{\Sigma}=\bm{I}$}:
\begin{equation}
    \dot{\bm{x}}(t) = \bm{\Sigma} \bm{\Sigma}^T \nabla \ln \rho_S(\bm{x}) + \sqrt{2}\bm{\Sigma} \bm{\xi}(t),
\end{equation}
where $\bm{\xi}(t)$ is a vector of independent delta-correlated Gaussian white noise processes. \added{Throughout all experiments, we fix $\bm{\Sigma}=\bm{I}$ (the identity matrix), which corresponds to isotropic diffusion. This choice is made for simplicity; a systematic procedure for constructing $\bm{\Sigma}$ from time-correlation functions in observational data is detailed in \cite{giorgini2025data, giorgini2025reduced}.} We evaluated the steady-state distributions of these generated trajectories and compared them with those obtained from the observed data to verify whether the KGMM-estimated score function correctly reproduces the invariant measure of the underlying dynamical system.

\added{For all systems except the KS equation, we use $N_{\text{eff}} = 10^5$ effectively uncorrelated samples for training. For the KS equation, data augmentation via circular shifts yields $N_{\text{eff}} = 8 \times 10^5$ effectively uncorrelated samples. The decorrelation time $t_d$ is estimated from the autocorrelation function of each coordinate as the first time at which the autocorrelation decays to $1/e$ of its initial value. Complete details, including $t_d$ for each system, are provided in \ref{app:hyperparameters} (Table 1).}

\added{The neural network architecture used for interpolation consists of fully connected layers with the Swish activation function \cite{ramachandran2017searching}, defined as $\varphi(x) = x \cdot \sigma(x)$ where $\sigma(x) = 1/(1+e^{-x})$ is the sigmoid function. Swish has been shown to outperform ReLU in various tasks due to its smoothness and non-monotonic behavior \cite{ramachandran2017searching}. For all the systems, we employed a three-layer architecture, with Swish activations between layers and a linear output layer. Training used the Adam optimizer \cite{kingma2014adam}. Complete hyperparameters (learning rates, batch sizes, epochs, $\sigma$, $N_C$, and sampling details) are listed in \ref{app:hyperparameters}.}

\subsection{Reduced Triad Model}
The triad model, as detailed in \citep{majda2009}, serves as a fundamental representation of nonlinear energy exchanges among interacting modes in turbulent systems. By leveraging timescale separation techniques, this system can be effectively reduced from its three-dimensional formulation to a one-dimensional stochastic differential equation, capturing the essential low-frequency behavior while parameterizing unresolved fast-scale interactions.

The resulting reduced-order stochastic differential equation takes the form:
\begin{equation}
\dot{x}(t) = F + a x(t) + b x^2(t) - c x^3(t) + \sigma_1\,\xi_1(t)+\sigma_2(x)\xi_2(t),
\label{reduced}
\end{equation}
where the deterministic drift coefficients and external forcing term are defined as:
\begin{equation}
\begin{aligned}
&a = -1.809, \quad
b = -0.0667, \quad
c = 0.1667, \\
&A = 0.1265, \quad B = -0.6325, \quad
F = \frac{A B}{2},
\end{aligned}
\end{equation}
and the noise amplitudes are given by:
\begin{equation}
\sigma_1 = 0.0632,\quad
\sigma_2(x) = A - Bx.
\end{equation}

An analytical expression for the score function of this model is available:
\begin{equation} 
s(x) = 2\frac{\frac{AB}{2} + (a - B^2) x + b x^2 - c x^3}{\sigma_1^2 + \sigma_2^2(x)},
\label{eq:reduced_score}
\end{equation}
where the denominator reflects the additive and multiplicative noise contributions. We used $\sigma$ in the range $[0.01, 0.05]$ with $N_C \approx 300$--$400$ (probability-threshold dependent) in Algorithm~\ref{algorithm}; \emph{the figure shown} was produced with $\sigma \approx 0.05$ and $N_C \approx 346$.

\begin{figure*}
 	\centering	 	\includegraphics[width=\textwidth]{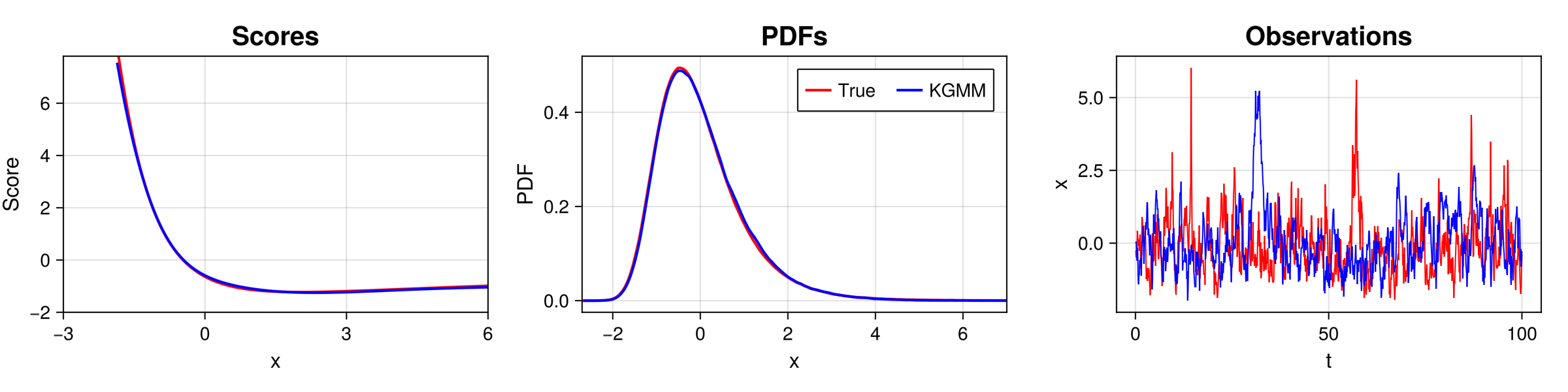}
 	\caption{Reduced triad model (Eq.~\eqref{reduced}). \textbf{Left panel:} Comparison between the KGMM-estimated score function and its analytical expression given by Eq.~\eqref{eq:reduced_score}. \textbf{Center panel:} Comparison between the observed steady-state distribution (True, red) and the one obtained from integrating Eq.~\eqref{eq:langevin} using the KGMM score function (KGMM, blue)\added{, demonstrating that KGMM correctly reproduces the invariant measure}. \textbf{Right panel:} Comparison between sample trajectories obtained by integrating Eq.~\eqref{reduced} (True, red) and Eq.~\eqref{eq:langevin} using the KGMM score function (KGMM, blue). \added{Note that individual trajectories differ due to stochastic realizations.}}
 	\label{fig1}
\end{figure*}

In Fig.~\ref{fig1} we compared the score function and the steady-state distribution estimated with the KGMM algorithm with their ground truths. As shown in the figure, the KGMM-estimated score function closely matches the analytical expression. Additionally, integrating Eq.~\eqref{eq:langevin} with the KGMM score function as the drift term successfully reconstructs the steady-state distribution and reproduces key statistical properties of the original system.

\subsection{Two-Dimensional Asymmetric Potential System}
The two-dimensional asymmetric potential system is governed by the stochastic differential equation:
\begin{equation}
\dot{\v{x}}(t) = -\nabla U(\v{x}) + \sqrt{2}\,\v{\xi}(t),
\label{lang}
\end{equation}
where the potential function \(U(\v{x})\) is given by:
\begin{equation}
U(\v{x}) = (x_1 + A_1)^2 (x_1 - A_1)^2 + (x_2 + A_2)^2 (x_2 - A_2)^2 + B_1 x_1 + B_2 x_2.
\label{potential_2D}
\end{equation}
The coefficients used in our study are:
\begin{equation}
A_1 = 1.0, \quad A_2 = 1.2, \quad B_1 = 0.6, \quad B_2 = 0.3.
\end{equation}

The corresponding score function is defined as:
\begin{equation}
\v{s}(\v{x}) = -\nabla U(\v{x}).
\end{equation}
We used $\sigma=0.05$ and $N_C=725$ inside Algorithm \ref{algorithm}.

This model describes an asymmetric potential landscape typical of systems exhibiting multistability, a feature often observed in climate models where multiple stable states can exist \cite{margazoglou2021dynamical}. The goal of our analysis is to compare the KGMM-estimated score function with the true score function and assess the accuracy of the reconstructed probability densities.

\begin{figure*}
 	\centering	\includegraphics[width=\textwidth]{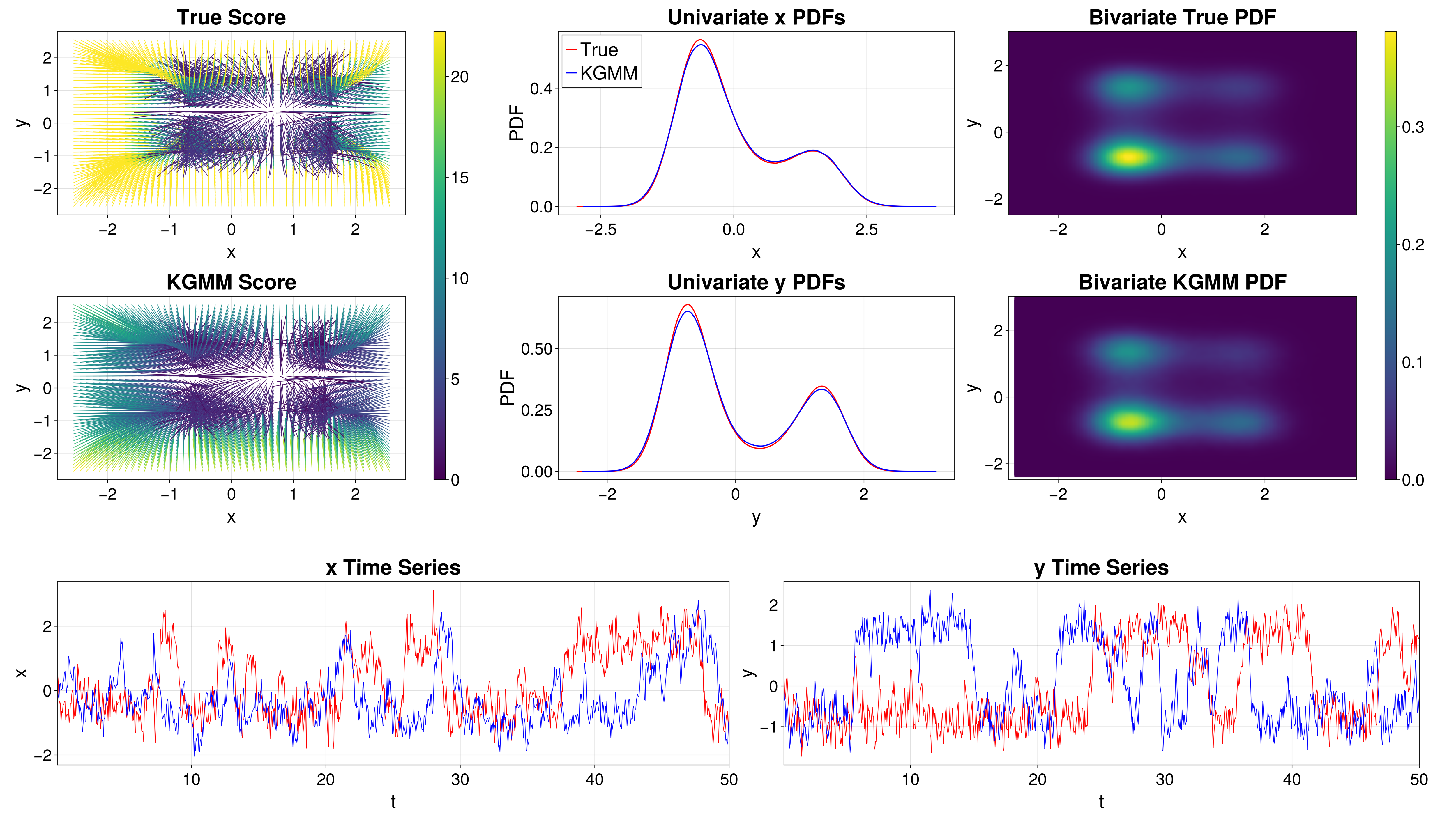}
 	\caption{\textbf{Two-dimensional asymmetric potential system.}
 	\textbf{First row, left:} The force field of the true score function (top) and the force field of the KGMM-estimated score function (bottom).
 	\textbf{First row, center:} Comparison between the observed univariate PDFs for $x$ (top) and $y$ (bottom) with those obtained by integrating Eq.~\eqref{lang} with the KGMM-estimated score function\added{, showing close agreement in marginal distributions}.
 	\textbf{First row, right:} Comparison between the observed bivariate probability density (top) and the reconstructed density using the KGMM-based score function (bottom)\added{, confirming reproduction of the joint distribution}.
    \textbf{Bottom row:} Comparison between sample trajectories for $x$ (left) and $y$ (right) obtained by integrating Eq.~\eqref{potential_2D} (True, red) and Eq.~\eqref{eq:langevin} using the KGMM score function (KGMM, blue). \added{Note that individual trajectories may differ due to stochastic realizations.}}
 	\label{fig2}
\end{figure*}

Figure~\ref{fig2} shows that the KGMM-estimated score function closely matches the analytical score function near the potential minima, where the majority of the observed data points are concentrated. Additionally, the probability density functions obtained using the KGMM-estimated score function agree well with those computed from direct observations.

However, discrepancies between the two score functions are observed in regions far from the potential minima. This deviation arises due to the scarcity of observed data points in these regions, leading to errors in the KGMM-based reconstruction of the score function.

\subsection{Stochastic Lorenz 63 Model}
The Lorenz 63 system \cite{lorenz1963} is a classical model for atmospheric convection, encapsulating key features of chaotic behavior in climate dynamics. Unlike the previous two models, the Lorenz 63 system is inherently chaotic. To capture the influence of unresolved processes occurring at shorter timescales, we consider a stochastic extension of the Lorenz 63 system by incorporating a noise term:
\begin{equation}\begin{split}
\dot{x}(t) &= \sigma (y(t) - x(t)) + \sigma_\xi\xi_1(t),\\
\dot{y}(t) &= x(t)(\rho - z(t)) - y(t) + \sigma_\xi\xi_2(t),\\
\dot{z}(t) &= x(t)y(t) - \beta z(t) + \sigma_\xi\xi_3(t),
\label{lorenz63}
\end{split}\end{equation}
where \(\xi_1(t), \xi_2(t),\) and \(\xi_3(t)\) are independent Gaussian white noise processes with unit variance. The coefficients used in our study are:
\begin{equation}
\sigma = 10.0, \quad \rho = 28.0, \quad \beta = \frac{8}{3}, \quad \sigma_\xi = 5.0.
\end{equation}
We used $\sigma=0.05$ and $N_C=754$ inside Algorithm \ref{algorithm}.

When comparing the trajectory of the original (chaotic) Lorenz 63 system with the trajectory obtained by integrating the corresponding Langevin equation \eqref{eq:langevin}
using the KGMM-estimated score function, the time evolution at short timescales can look qualitatively very different. This occurs because the deterministic details in the original chaotic system generate specific trajectories that are highly sensitive to initial conditions, whereas the Langevin approach encodes the steady-state behavior through noise-driven dynamics and does not preserve the exact local chaotic structure. Nevertheless, on longer timescales, the two systems share the same invariant measure, as the KGMM score function accurately reproduces the statistical properties observed in the data.

\begin{figure*}
 	\centering	\includegraphics[width=\textwidth]{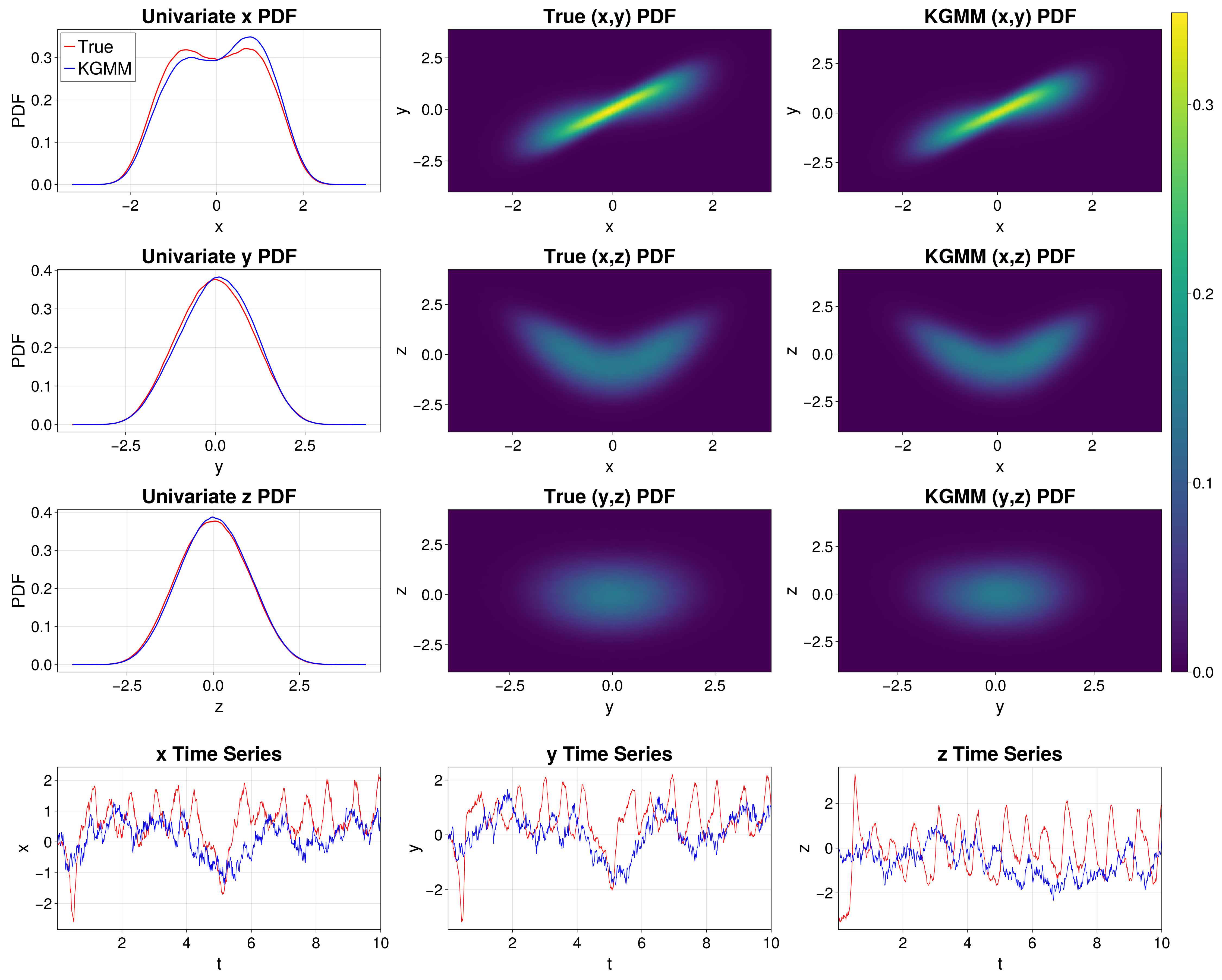}
 	\caption{\textbf{Lorenz 63 system.}
 	\textbf{First column:} Comparison between the observed univariate PDFs for $x$, $y$, and $z$ (True, red) and those obtained integrating the Langevin equation using the KGMM-estimated score function (KGMM, blue)\added{, demonstrating accurate marginal distributions}.
 	\textbf{Second and third columns:} Comparison between the observed bivariate PDFs for $(x,y)$, $(x,z)$, and $(y,z)$ (True, left column) and those obtained using the KGMM-based score function (KGMM, right column)\added{, showing faithful reproduction of joint statistics despite different short-time trajectory behavior}.
    \textbf{Bottom row:} Comparison between sample trajectories for $x$, $y$, and $z$ obtained by integrating Eq.~\eqref{lorenz63} (True, red) and Eq.~\eqref{eq:langevin} using the KGMM score function (KGMM, blue). \added{Note that individual trajectories may differ due to stochastic realizations.}}
 	\label{fig3}
\end{figure*}

As shown in Fig.~\ref{fig3}, the KGMM-estimated score function successfully reconstructs the steady-state probability distributions of the system. Despite the short-timescale trajectory differences, the long-term statistical agreement demonstrates the robustness of the KGMM approach in capturing the essential invariant features of a chaotic system.

\subsection{Stochastic Lorenz 96 Model}
The Lorenz 96 model \citep{lorenz1996}, is a paradigmatic system for studying multiscale chaotic dynamics, originally designed as a simplified model of atmospheric circulation. It consists of a set of slow variables, \(x_k\), which evolve on a longer timescale, coupled to a set of fast variables, \(y_{k,j}\), representing small-scale turbulent fluctuations. To account for unresolved processes occurring on timescales even shorter than those explicitly modeled, we consider a stochastic extension of the system:
\begin{equation}
\frac{\d x_k}{\d t} = -x_{k-1}(x_{k-2} - x_{k+1}) - \nu x_k + F + c_1 \sum_{j=1}^{N_j} y_{k,j} + \sigma \xi_k(t),
\end{equation}
\begin{equation}
\frac{\d y_{k,j}}{\d t} = -c b y_{k,j+1} (y_{k,j+2} - y_{k,j-1}) - c \nu y_{k,j} + c_1 x_k+ \sigma \xi_{k,j}(t).
\label{lorenz96}
\end{equation}
Here, \(\xi_k(t),\xi_{k,j}(t)\) are uncorrelated Gaussian white noise processes with unit variance, representing the effect of high-frequency fluctuations not explicitly resolved. The model parameters are chosen as follows:
\begin{equation}\begin{split}
&F = 4.0, \quad \nu = 1.0, \quad c = 10.0, \\ & b = 10.0, \quad c_1 = \frac{c}{b} = 1.0, \quad \sigma = 0.2.
\end{split}\end{equation}
This formulation naturally introduces three distinct timescales into the system. The shortest timescale is associated with the stochastic forcing term, the intermediate timescale corresponds to the chaotic dynamics of the 40-dimensional fast process \(\{y_{k,j}\}\), and the longest timescale governs the evolution of the 4-dimensional slow variables \(\{x_k\}\). We used $\sigma=0.05$ and $N_C=3818$ inside Algorithm \ref{algorithm}.

Similar to the Lorenz 63 case, comparing the short-timescale behavior of the original Lorenz 96 trajectories with those obtained by integrating \eqref{eq:langevin} using the KGMM-estimated score function reveals qualitative differences due to the deterministic chaotic nature of the full Lorenz 96 model. However, as time evolves, both the original system and the KGMM-based Langevin model settle into the same statistical regime, sharing the same invariant measure. Due to the symmetries in the system, we present only the trajectory and univariate distribution for a single variable, as the behavior of the remaining variables is statistically equivalent.

\begin{figure*}
 	\centering	 	\includegraphics[width=\textwidth]{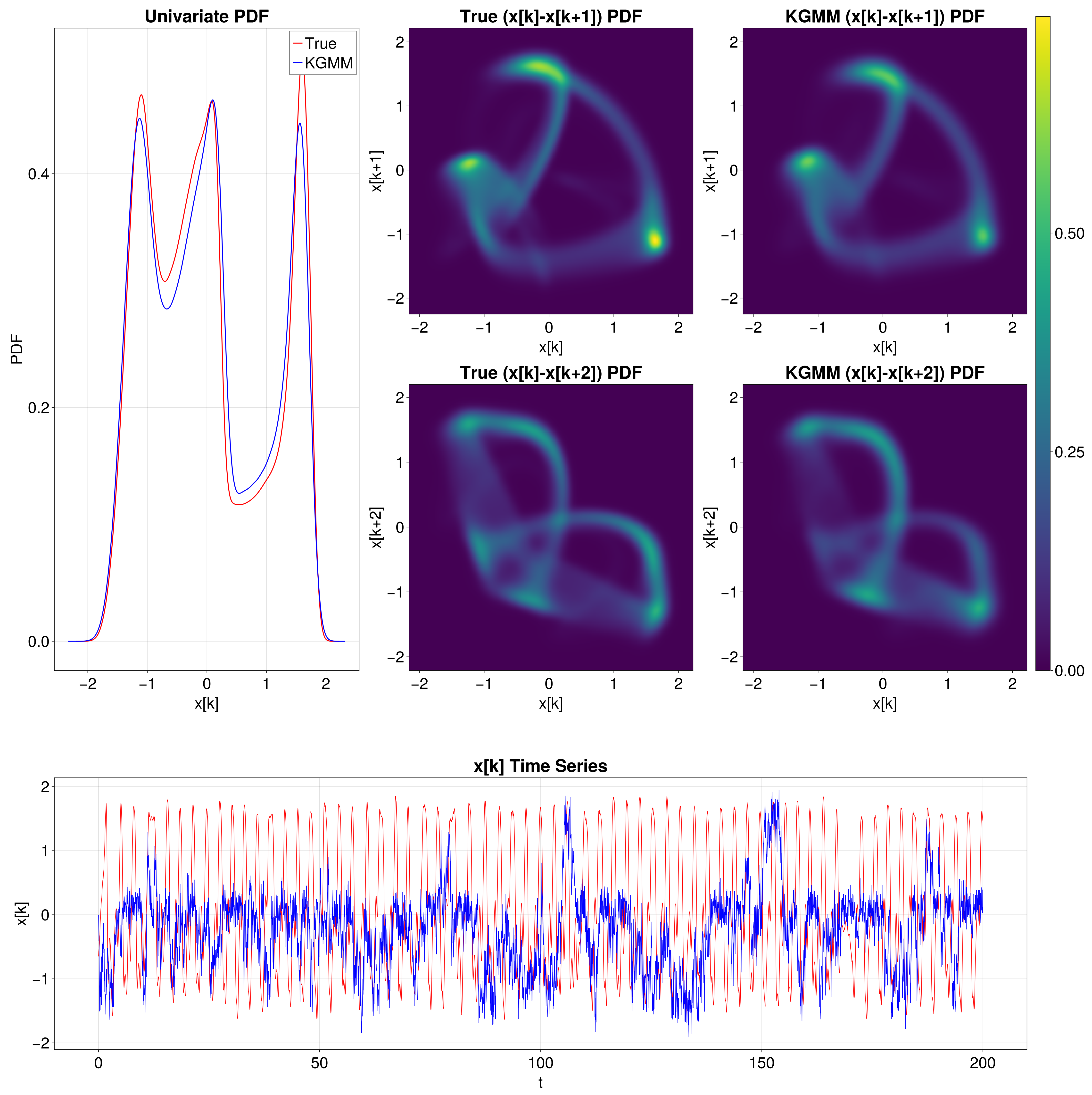}
 	\caption{\textbf{Lorenz 96 system.} 
 	\textbf{Top left:} Comparison between the observed univariate PDF (True, red) and the one obtained integrating the Langevin equation using the KGMM-estimated score function (KGMM, blue)\added{, showing excellent agreement}.
 	\textbf{Top center and right:} Comparison between the observed bivariate PDFs for $x[k]$-$x[k+1]$ (center) and $x[k]$-$x[k+2]$ (right) (True, top row) and those obtained using the KGMM-based score function (KGMM, bottom row)\added{, demonstrating that KGMM captures the non-Gaussian structure of the invariant measure}.
    \textbf{Bottom panel:} Comparison between sample trajectories for $x[k]$ obtained by integrating Eq.~\eqref{lorenz96} (True, red) and Eq.~\eqref{eq:langevin} using the KGMM score function (KGMM, blue). \added{Note that individual trajectories may differ due to stochastic realizations.}}
 	\label{fig4}
\end{figure*}

The degree of chaos in the Lorenz 96 system depends on the magnitude of the external forcing \(F\) and the number of slow variables \(N_k\). For larger values of \(F\) and \(N_k\), the system exhibits fully developed turbulence, and its steady-state distribution approaches a multivariate Gaussian. In this study, we focus on an intermediate chaotic regime where the steady-state PDF deviates significantly from a Gaussian distribution. This choice allows us to better assess the ability of the KGMM method to accurately reconstruct non-Gaussian statistical structures, which would be harder to detect in a system where the steady-state distribution is trivially Gaussian.

\subsection{\added{Kuramoto--Sivashinsky Equation}}

\added{The Kuramoto--Sivashinsky (KS) equation is a prototypical model for spatiotemporal chaos arising in pattern formation, flame-front dynamics, and fluid instabilities \cite{kuramoto1976persistent, sivashinsky1977nonlinear}. The one-dimensional KS equation on a periodic domain is given by}
\begin{equation}\added{
\frac{\partial u}{\partial t} = - \Delta u - \Delta^2 u - \frac{1}{2} |\nabla u|^2,
\label{eq:ks}
}\end{equation}
\added{where $u(x,t)$ is a scalar field, $\Delta = \partial^2/\partial x^2$ is the Laplacian, $\nabla = \partial/\partial x$ is the spatial derivative, and the nonlinear term $|\nabla u|^2$ represents advection. The domain size, $L=34$, is the control parameter that transitions the system to chaotic dynamics.  The KS equation exhibits high-dimensional chaotic attractors and has been extensively studied as a benchmark for reduced-order modeling and data-driven methods \cite{cvitanovic2010chaos, lu2019prediction, giorgini2024reduced}.}

\added{We apply KGMM to finite-dimensional projections of the KS attractor obtained from the same underlying dataset with $n_{\text{grid}}=128$ Fourier modes. By subsampling with different spatial stride values $n_{\text{stride}} \in \{32,16,8\}$, we obtain reduced state vectors of dimensions $d \in \{4,8,16\}$, respectively. To ensure sufficient training data for the higher-dimensional cases---where the required number of clusters approaches the total number of uncorrelated samples---we adopted a denser temporal sampling strategy. Specifically, instead of extracting one snapshot per decorrelation time $t_d$ (as done for the other systems), we sampled one snapshot every $t_d/10$ from the KS time series. Data augmentation via circular shifts applied to each snapshot produces 8 uncorrelated realizations per snapshot, yielding $N_{\text{eff}} = 8 \times 10^5$ effectively uncorrelated samples. The centered and normalized mode amplitudes are then used to train the KGMM score estimator with $\sigma = 0.1$ and cluster counts $N_C = 74{,}047$ ($d=4$), $N_C = 747{,}507$ ($d=8$), and $N_C = 1{,}297{,}386$ ($d=16$). The neural network architecture consists of two hidden layers with $[128, 64]$ neurons, Swish activations, and a linear output layer.}

\begin{figure*}\added{
 	\centering	\includegraphics[width=\textwidth]{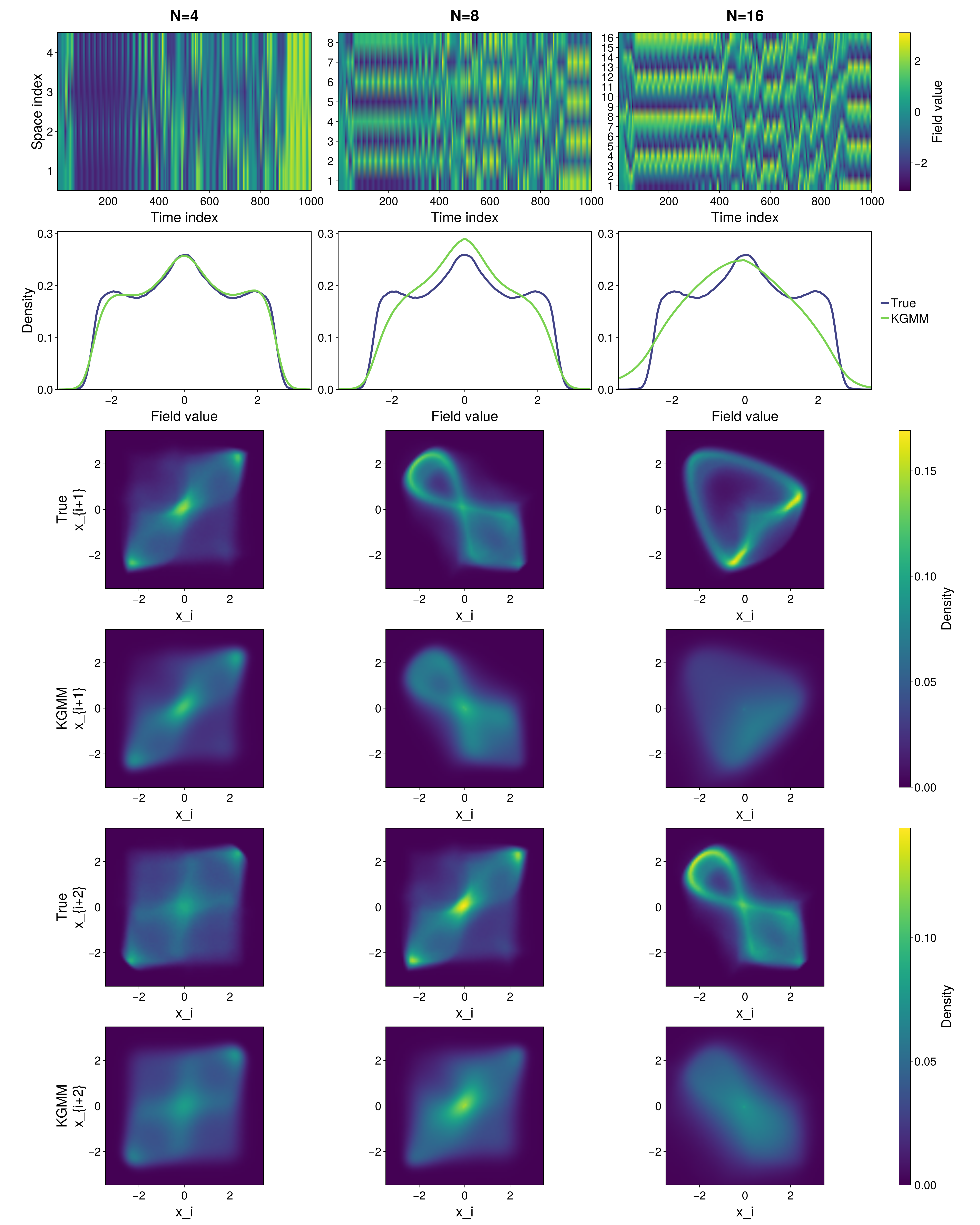}
 	\caption{\textbf{Kuramoto--Sivashinsky equation in reduced coordinates.} Comparison of true statistics (from full KS simulation, subsampled) and KGMM-generated statistics for three dimensional cases: $d=4$ (left column), $d=8$ (middle column), and $d=16$ (right column). \textbf{Row 1:} Spatiotemporal evolution (space index vs. time). \textbf{Row 2:} Averaged univariate PDF (True vs. KGMM). \textbf{Rows 3--4:} Averaged bivariate PDF at distance 1 (True above, KGMM below). \textbf{Rows 5--6:} Averaged bivariate PDF at distance 2.  Colorbars are shared within each set of bivariate plots.}
 	\label{fig:ks}
}\end{figure*}

\added{Figure~\ref{fig:ks} presents a comprehensive comparison between the true KS dynamics (subsampled and centered) and the KGMM-generated statistics for all three dimensional cases. The top row shows spatiotemporal plots of the subsampled KS field over time (space index vs. time index) obtained by direct integration of the KS equation. The second row shows averaged univariate PDFs obtained by marginalizing over all spatial modes, comparing the empirical distribution (True) with the KGMM-generated distribution. Rows 3--4 display averaged bivariate PDFs for spatial correlations at distance 1 (adjacent modes), with the true joint distribution shown in row 3 and the KGMM-reconstructed distribution in row 4. Similarly, rows 5--6 present averaged bivariate PDFs for spatial correlations at distance 2. The colormap is shared across corresponding bivariate panels to facilitate comparison.}

\added{The figure reveals a decrease in PDF reconstruction performance as the dimension increases from $d=4$ to $d=16$. This degradation arises because the number of clusters needed to accurately reconstruct the score function grows exponentially with the effective dimension of the system. For $d=8$ and $d=16$, the cluster counts ($N_C = 790{,}637$ and $N_C = 1{,}297{,}386$, respectively) approach the number of effectively uncorrelated data points available. In this regime, KGMM offers limited computational advantage, since we cannot substantially reduce the number of training points for the neural network compared to plain DSM. Consequently, the method incurs the computational overhead of clustering without fully realizing the efficiency gains that make KGMM attractive for lower-dimensional problems. For high-dimensional systems, KGMM becomes beneficial only when the dataset size far exceeds the requisite number of clusters. In practice, such large datasets are often unavailable, making plain DSM with appropriately designed, physics-informed neural network architectures a more practical choice for very high-dimensional systems.}

\added{These results also highlight the critical role that attractor dimensionality plays in determining the required number of clusters. Comparing the KS equation at $d=4$ with the Lorenz~96 system (also $d=4$), we observe that achieving comparable reconstruction accuracy for KS required approximately 20 times more clusters ($N_C = 74{,}047$ versus $N_C = 3{,}818$), despite both systems residing in the same ambient dimension. This disparity arises from differences in the intrinsic dimensionality of the respective attractors. Examination of the bivariate PDFs reveals that the KS distribution occupies a substantially larger fraction of the state space: its support extends over a genuinely two-dimensional region, whereas the Lorenz~96 distribution is concentrated along a lower-dimensional manifold---a narrow, elongated subset of the plane. Geometrically, the KS attractor exhibits higher effective dimension, meaning that the invariant measure is distributed across a larger set in phase space. Consequently, partitioning the support of the KS distribution into regions of comparable local homogeneity demands a finer tessellation, and hence a larger number of clusters, to adequately resolve the spatial structure of the score function. This observation underscores that the computational cost of KGMM is governed not merely by the nominal dimension $d$, but more fundamentally by the intrinsic dimension of the attractor and the geometric complexity of the invariant measure's support.}

\section{\added{Performance Comparison and Limitations}}
\added{In this section, we compare the computational performance of training score estimators using KGMM preprocessing versus direct neural network training on the full dataset (standard DSM). We also discuss practical guidelines for selecting hyperparameters, particularly $N_C$ and $\sigma$, and address the limitations of the KGMM approach.}

\subsection{\added{Computational Performance: KGMM vs. Direct Training}}
\added{To understand when KGMM offers computational advantages over direct DSM training, we analyze the time complexity of both approaches. For direct DSM training on a dataset of $N$ points, the neural network processes all $N$ samples in each epoch. The total cost for $n_{\text{epochs}}$ epochs scales as}
\begin{equation}\added{
T_{\text{direct}} = \mathcal{O}(n_{\text{epochs}} \cdot N \cdot D \cdot H),
}\end{equation}
\added{where $D$ is the state-space dimension and $H$ represents the network complexity (proportional to the total number of parameters).}

\added{In contrast, KGMM consists of two sequential phases: preprocessing and neural network training. The preprocessing phase performs bisecting K-means clustering to partition the $N$ data points into $N_C$ clusters. The bisecting strategy achieves $\mathcal{O}(N \cdot D \cdot \log N_C)$ complexity per iteration through hierarchical binary splits. Following clustering, we iteratively refine the cluster-wise score estimates via an exponential moving average (EMA) procedure. Each EMA iteration assigns perturbed points $\bm{x}_i = \bm{\mu}_i + \bm{z}_i$ to their nearest cluster centroids (costing $\mathcal{O}(N \cdot D \cdot \log N_C)$ using efficient tree-based search) and updates the running average of $\bm{z}_i$ within each cluster. If we denote by $i_{\text{EMA}}$ the number of EMA iterations required for convergence (typically $i_{\text{EMA}} \in [5, 10]$ in our experiments), the preprocessing phase has total complexity}
\added{\begin{equation}\begin{split}
T_{\text{preprocess}} &= \mathcal{O}(N \cdot D \cdot \log N_C) + \mathcal{O}(i_{\text{EMA}} \cdot N \cdot D \cdot \log N_C) \\ &= \mathcal{O}(i_{\text{EMA}} \cdot N \cdot D \cdot \log N_C).
\end{split}\end{equation}}
\added{Subsequently, the neural network is trained on only $N_C$ cluster centroids (rather than all $N$ data points), yielding training cost}
\begin{equation}\added{
T_{\text{train}} = \mathcal{O}(n_{\text{epochs}} \cdot N_C \cdot D \cdot H).
}\end{equation}
\added{The total KGMM cost is thus}
\added{\begin{equation}\begin{split}
T_{\text{KGMM}} &= T_{\text{preprocess}} + T_{\text{train}} \\ &= \mathcal{O}(i_{\text{EMA}} \cdot N \cdot D \cdot \log N_C) + \mathcal{O}(n_{\text{epochs}} \cdot N_C \cdot D \cdot H).
\end{split}\end{equation}}
\added{KGMM becomes computationally advantageous when $T_{\text{KGMM}} < T_{\text{direct}}$. For typical values where the network complexity dominates ($H \gg \log N_C$) and convergence requires moderate iteration counts ($i_{\text{EMA}} \sim 5$--$10$, $n_{\text{epochs}} \sim 100$--$1000$), this condition simplifies to requiring $N_C \ll N$. In the regime where the cluster count is one or two orders of magnitude smaller than the dataset size ($N_C/N \in [0.01, 0.1]$), the amortized preprocessing cost is substantially outweighed by the savings from training on $N_C$ rather than $N$ points.}

\added{Both KGMM and plain DSM benefit from data-parallel and GPU-accelerated implementations. In KGMM, the bisecting K-means assignments and distance computations across all points, as well as the EMA updates of cluster statistics, are embarrassingly parallel operations over the dataset and clusters; they map naturally to SIMD/SIMT kernels and can be distributed across multiple devices. Likewise, the subsequent neural-network training (both for KGMM interpolation and for plain DSM) proceeds via mini-batch stochastic optimization, which supports efficient batching on GPUs and multi-GPU data parallelism. In practice, keeping data resident on device and vectorizing nearest-centroid queries and reductions yields near-linear scaling with hardware throughput.}

\added{We validate these expectations on two low-dimensional systems (Reduced Triad and Lorenz~63) by training score estimators with and without KGMM preprocessing. For plain DSM, we vary the number of training epochs to explore the accuracy-time trade-off; for KGMM, we vary the number of clusters $N_C$. The choice to vary $N_C$ rather than epochs for KGMM reflects the fact that the clustering and EMA iterations constitute the dominant computational bottleneck in the KGMM pipeline. Since each EMA iteration must assign all $N$ perturbed points to their nearest cluster centroids and update cluster statistics, this preprocessing phase scales with $N$ and typically consumes most of the total wall-clock time, whereas the subsequent neural network training on only $N_C$ centroids is comparatively fast. Thus, varying $N_C$ directly controls the primary source of computational cost in KGMM.}

\begin{figure*}\added{
	\centering	\includegraphics[width=\textwidth]{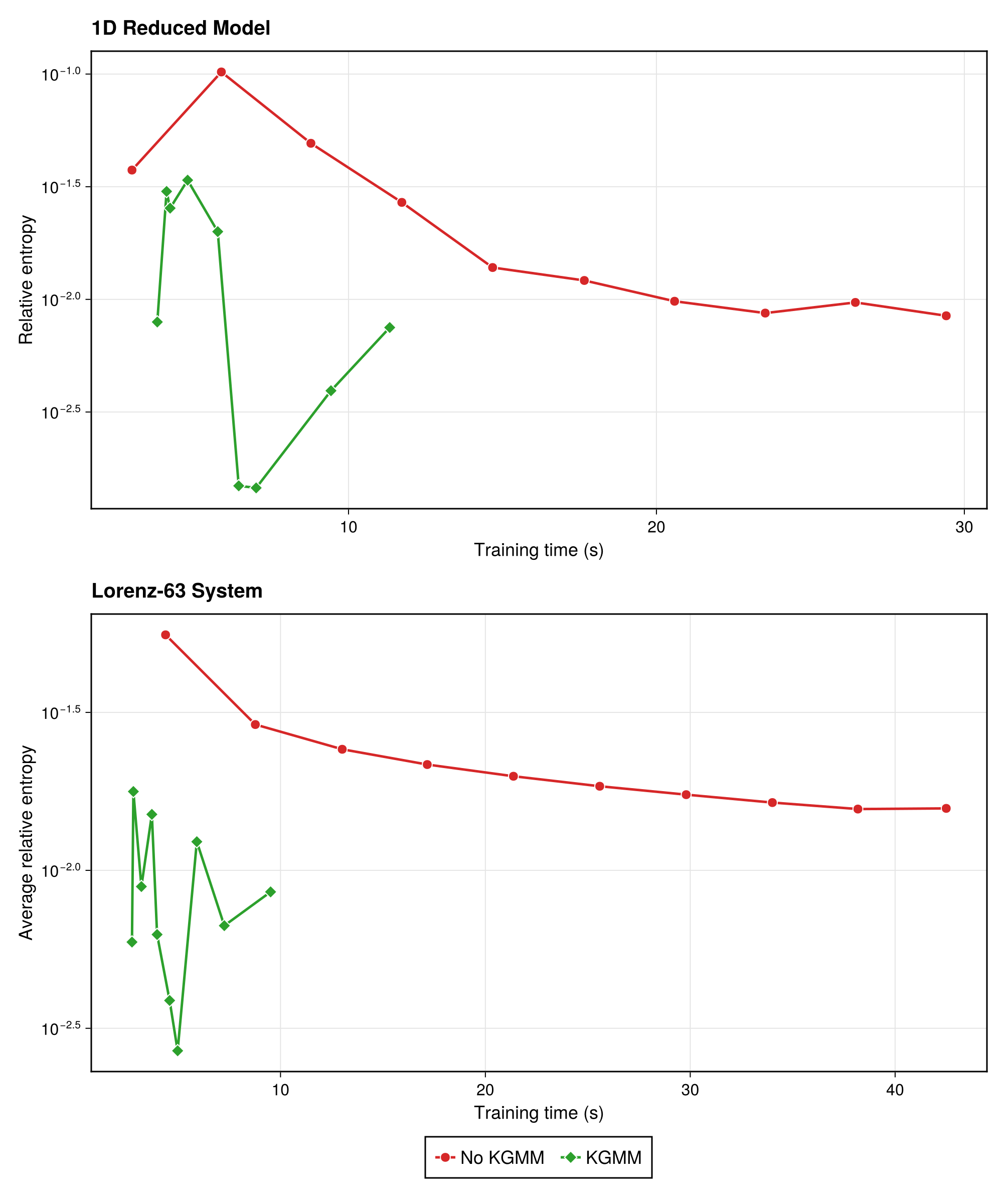}
	\caption{\textbf{Performance comparison: KGMM vs. direct training.} Each row shows relative entropy (KL divergence) versus total computational time. \textbf{Top:} Reduced Triad. \textbf{Bottom:} Lorenz 63. Direct training (red) exhibits a generally monotonic decrease of relative entropy with time, whereas KGMM (green) displays an optimal regime in $N_C$ beyond which performance worsens due to increased per-cluster noise. KGMM achieves significantly better accuracy at substantially reduced computational cost, demonstrating order-of-magnitude speedups in the optimal regime.}
	\label{fig:performance}
}\end{figure*}

\added{We measure performance using the relative entropy (Kullback--Leibler divergence) $D_{\text{KL}}(\rho_{\text{true}} \| \rho_{\text{est}})$ between the true stationary distribution and the distribution generated by integrating the Langevin equation with the estimated score function; for Lorenz~63 we report the average of the KL divergences of the three univariate marginals ($x$, $y$, $z$). Figure~\ref{fig:performance} plots relative entropy versus total computational time (wall-clock seconds) for both methods. For direct DSM training (red curves), the relative entropy decreases monotonically with computational time, modulo stochastic fluctuations, as more training epochs refine the neural network approximation. In contrast, KGMM (green curves) exhibits a qualitatively different behavior: there exists an optimal cluster count $N_C^*$ that minimizes the relative entropy. Below this optimum, increasing $N_C$ improves the geometric resolution of the score function by placing cluster centroids closer together, enabling the neural network to interpolate more accurately. However, beyond $N_C^*$, further increases in cluster count reduce the number of data points per cluster, amplifying statistical noise in the cluster-wise score estimates $\bm{q}_k$, which degrades accuracy despite finer spatial resolution.}

\added{Crucially, Figure~\ref{fig:performance} demonstrates that KGMM achieves substantially lower relative entropy at significantly reduced computational cost compared to direct training. This efficiency gain arises because KGMM provides statistically precise score estimates at each cluster centroid by averaging noise vectors $\bm{z}_i$ over all data points assigned to that cluster. Consequently, the neural network trains on a dataset of size $N_C \ll N$ consisting of high-quality, low-noise target values, rather than on $N$ individual noisy samples as in standard DSM. This dual advantage—fewer training points and higher-quality targets—accounts for both the reduced training time and the superior accuracy of KGMM in the optimal regime.}

\subsection{\added{Hyperparameter Selection and Discussion}}
\added{The KGMM method introduces two primary hyperparameters: the noise level $\sigma$ and the number of clusters $N_C$. We now discuss practical strategies for choosing these parameters and acknowledge the limitations of our approach.}

\subsubsection{\added{Choice of $\sigma$}}
\added{The noise level $\sigma$ controls the smoothness of the estimated score function. In theory, $\sigma \to 0$ recovers the true score $\nabla \log \rho_S$, but in practice, finite-$\sigma$ bias and finite-sample noise must be balanced. Smaller $\sigma$ yields more accurate approximations of $\rho_S$ but amplifies noise in regions of low data density, whereas larger $\sigma$ provides smoother estimates at the cost of blurring fine-scale structure.}

\added{In our experiments, we found that $\sigma \in [0.01, 0.1]$ (in normalized coordinates) works well for a wide range of systems. A heuristic rule is to choose $\sigma$ proportional to the typical inter-sample distance in regions of moderate density: $\sigma \approx c \cdot (\text{characteristic length scale})$, where $c \in [0.1, 0.5]$. For the systems studied here, we used $\sigma = 0.01$ for the Reduced Triad, $\sigma = 0.05$ for the 2D Potential and Lorenz systems, and $\sigma = 0.1$ for the KS equation. As a practical rule of thumb, set $\sigma = 0.05$ by default and increase it to $\sigma = 0.1$ if the average number of data points per cluster falls below 10. We did not rigorously optimize $\sigma$ in this work, leaving systematic hyperparameter tuning for future investigation.}

\subsubsection{\added{Choice of $N_C$}}
\added{The number of clusters $N_C$ determines the spatial resolution of the score function estimates. Larger $N_C$ improves resolution but reduces the number of samples per cluster, increasing statistical noise. Conversely, smaller $N_C$ oversmooths the score function, particularly in regions of rapid gradient variation.}

\added{Our experiments across multiple systems reveal a consistent empirical relationship between the optimal cluster count and the hyperparameters:}
\begin{equation}\added{
N_C \propto \sigma^{-d_{\text{eff}}},
\label{eq:nc_scaling}
}\end{equation}
\added{where $d_{\text{eff}}$ denotes the effective dimension of the attractor (the intrinsic dimensionality of the support of $\rho_S$, which may be smaller than the ambient state-space dimension for systems with strong dimensional reduction). This scaling relationship follows from geometric considerations: the noise level $\sigma$ defines a characteristic length scale over which the score function is smoothed by the Gaussian convolution. To resolve spatial variations in $\nabla \log p_\sigma(\bm{x})$, cluster centroids must be spaced at intervals comparable to $\sigma$. In a $d_{\text{eff}}$-dimensional space, covering the attractor with such clusters requires $N_C \sim (\text{diameter}/\sigma)^{d_{\text{eff}}} \propto \sigma^{-d_{\text{eff}}}$.}

\added{In practice, the optimal $N_C$ also depends on the available sample size $N$: when $N_C$ approaches $N$, each cluster contains too few points for reliable averaging, degrading performance. We find that the regime $N_C / N \in [0.01, 0.1]$ balances geometric resolution with statistical reliability across the systems tested. The finite-$\sigma$ bias inherent in KGMM (analogous to that in DSM \cite{song2021}) implies that even with large $N_C$, the recovered score function approximates $\nabla \log p_\sigma$ rather than $\nabla \log \rho_S$ exactly. This bias can be partially mitigated by using smaller $\sigma$, at the cost of increased sensitivity to noise and the need for correspondingly larger $N_C$ per Eq.~\eqref{eq:nc_scaling}.}

\subsubsection{\added{Special Case: Gaussian Distributions}}
\added{It is instructive to consider the special case where the true stationary distribution $\rho_S$ is exactly Gaussian, $\rho_S(\bm{x}) = \mathcal{N}(\bm{x} \mid \bm{\mu}_{\text{true}}, \bm{\Sigma}_{\text{true}})$, for which the score function has the simple analytical form $\nabla \log \rho_S(\bm{x}) = -\bm{\Sigma}_{\text{true}}^{-1}(\bm{x} - \bm{\mu}_{\text{true}})$. A reader familiar with GMMs might initially assume that a single Gaussian component ($N_C = 1$) should suffice to recover this linear score function. However, this intuition is misleading in the context of KGMM due to the nature of the approximation.}

\added{Recall that in the GMM formulation underlying KGMM, we model the density as $p_\sigma(\bm{x}) = \sum_{k=1}^{K} w_k \mathcal{N}(\bm{x} \mid \bm{\mu}_k, \sigma^2 \bm{I})$, where we take $K = N$ (the number of data points) with each $\bm{\mu}_k$ equal to a data point and $\sigma^2 \bm{I}$ an isotropic covariance. This mixture of narrow Gaussians centered at the data points is fundamentally different from the true Gaussian distribution $\rho_S$ with covariance $\bm{\Sigma}_{\text{true}}$, even when $\rho_S$ itself is Gaussian. The convolution $p_\sigma(\bm{x}) = \int \rho_S(\bm{\mu}) \mathcal{N}(\bm{x} \mid \bm{\mu}, \sigma^2 \bm{I}) \d\bm{\mu}$ yields a broadened Gaussian with covariance $\bm{\Sigma}_{\text{true}} + \sigma^2 \bm{I}$, whose score differs from that of $\rho_S$.}

\subsubsection{\added{Advantages and Disadvantages of KGMM}}
\added{To summarize, KGMM offers several advantages:}
\begin{itemize}\added{
\item \textbf{Computational efficiency:} By clustering the data into $N_C \ll N$ representative points and training the neural network on these cluster centroids rather than all $N$ data points, KGMM reduces the number of training samples by one to two orders of magnitude. As demonstrated in Figure~\ref{fig:performance}, this reduction yields substantial computational savings, with KGMM achieving large speedups over direct DSM training while maintaining or improving accuracy.
\item \textbf{Robustness to small $\sigma$:} Unlike standard GMM approaches that compute the score function by explicitly differentiating the mixture density—a process that amplifies noise as $\sigma \to 0$—KGMM avoids differentiation entirely. Instead, it directly estimates the score by averaging noise vectors $\bm{z}_i$ within each cluster, a statistically stable operation that mitigates noise amplification even for small covariance amplitudes (see Figure~\ref{fig:score_estimate}).
\item \textbf{Interpretability and statistical guarantees:} The cluster-wise score estimates $\bm{q}_k$ provide direct, interpretable insight into the local gradient structure at each centroid. For sufficiently large $N$ and $N_C$ chosen to adequately resolve the geometric structure of the score function, the law of large numbers guarantees that $\bm{q}_k$ converges to the true conditional expectation $\mathbb{E}[\bm{z} \mid \bm{x} \in \Omega_k]$ at each cluster centroid. Crucially, the neural network in KGMM serves solely as an interpolant between these statistically precise estimates; it is trained to fit the cluster centroids exactly, and overfitting to these target values is not only acceptable but desirable, as it ensures faithful reproduction of the preprocessed score estimates. In contrast, plain DSM requires the neural network to simultaneously learn the score structure and perform implicit regularization of noisy training samples. In that setting, overfitting to individual data points degrades generalization, necessitating careful regularization strategies. By decoupling statistical estimation (via clustering) from interpolation (via neural network fitting), KGMM circumvents this tension and provides well-defined target values that the network should reproduce without concern for overfitting.
\item \textbf{Flexibility:} KGMM can be combined with any neural network architecture for interpolation, making it modular and extensible.
}\end{itemize}
\added{However, KGMM also has limitations:}
\begin{itemize}\added{
\item \textbf{Hyperparameter sensitivity:} The performance depends on the choice of $N_C$ and $\sigma$, which currently lack rigorous tuning rules (though the heuristics in Eq.~\eqref{eq:nc_scaling} and the validation strategies above provide practical guidance).
\item \textbf{Finite-$\sigma$ bias:} Like DSM, KGMM learns the score of the convolved distribution $p_\sigma$ rather than the true $\rho_S$, introducing smoothing for finite $\sigma$.
\item \textbf{Curse of dimensionality:} The cluster count $N_C \propto \sigma^{-d}$ grows exponentially with dimension $d$, limiting scalability to very high dimensions ($d > 20$) unless combined with dimensionality reduction or manifold learning.
}
\end{itemize}
\added{Despite these limitations, KGMM provides a practical and efficient framework for score estimation in systems with moderate dimensionality ($d \lesssim 10$) and large sample sizes ($N \gtrsim 10^4$), as demonstrated by the results in Section 3.}

\section{Conclusions}
We have presented a hybrid method for estimating the score function by leveraging Gaussian Mixture Models and bisecting K-means clustering (KGMM). Our approach overcomes the noise amplification issues encountered in direct GMM-based methods for small covariance amplitudes and efficiently recovers the long-term statistical properties of both low-dimensional potential systems and chaotic Lorenz-type models. \added{We have demonstrated the scalability of KGMM to moderately high-dimensional systems by applying it to the Kuramoto--Sivashinsky equation in dimensions up to 16, confirming that the method preserves univariate and bivariate statistical structure even in the presence of spatiotemporal chaos.} Although the resultant stochastic trajectories may differ in their short-timescale details from those of the original chaotic systems, they converge to the same invariant measures, indicating that KGMM accurately reproduces the essential large-timescale dynamics.

\added{We have also compared the computational performance of KGMM preprocessing against direct neural network training on two low-dimensional test cases (Reduced Triad and Lorenz 63). Our results demonstrate that KGMM achieves substantially lower relative entropy at significantly reduced computational cost compared to standard DSM. This efficiency gain arises from two complementary mechanisms: (i) the neural network trains on only $N_C \ll N$ cluster centroids rather than all $N$ data points, reducing the training burden, and (ii) the cluster-wise score estimates are statistically precise due to averaging over many samples per cluster, providing high-quality training targets that enable faster convergence. The method's relation to Denoising Score Matching has been clarified, highlighting that both approaches share a finite-$\sigma$ bias but differ in their computational strategies: KGMM leverages explicit clustering and statistical estimation before neural network interpolation, whereas DSM trains end-to-end on the full dataset.}

\added{We have discussed practical guidelines for hyperparameter selection, including heuristic scaling rules for the number of clusters ($N_C \propto \sigma^{-d}$) and validation strategies for choosing the noise level $\sigma$. We have also acknowledged the limitations of KGMM, including its sensitivity to hyperparameters, finite-$\sigma$ bias, and exponential scaling of $N_C$ with dimension, which limits applicability to very high-dimensional systems without dimensionality reduction.}

\added{Beyond methodological developments, KGMM has demonstrated its versatility across multiple application domains. The algorithm has been successfully employed to estimate system responses via the generalized fluctuation-dissipation theorem \cite{giorgini2025predicting}, to construct data-driven reduced-order models from high-dimensional simulations \cite{giorgini2025data}, and to perform statistical parameter calibration in stochastic dynamical systems \cite{giorgini2025statistical}. Notably, the supplementary material of \cite{giorgini2025predicting} demonstrates KGMM performance on systems in dimensions 1--6 using an order of magnitude fewer samples than employed in the present work, showing that the method remains robust even with significantly reduced data. These applications highlight the broad utility of accurate score function estimation and underscore the practical impact of KGMM in enabling data-driven inference for complex systems.}

\added{The KGMM algorithm exhibits an inherently parallel structure that makes it particularly well-suited for GPU acceleration. Both the clustering phase (bisecting K-means with iterative centroid assignment) and the EMA iteration loop (assigning perturbed points to clusters and updating statistics) consist of embarrassingly parallel operations over the dataset. Future work will focus on developing a GPU-parallelized implementation to fully exploit this scalability, which will yield substantial performance improvements for large-scale datasets. Additionally, we plan to combine KGMM with dimensionality reduction techniques such as variational autoencoders to address very high-dimensional systems ($d > 10$). In this framework, an autoencoder would first map the high-dimensional state space to a lower-dimensional latent representation, KGMM would then estimate the score function in the latent space, and the learned score could be lifted back to the original coordinates via the decoder. This hybrid approach would leverage the curse-of-dimensionality mitigation provided by autoencoders while retaining the statistical robustness and computational efficiency of KGMM in the reduced latent space.} \added{Another promising direction is the development of adaptive methods for selecting $\sigma$ and $N_C$ during training, potentially using multi-scale or annealing strategies.}

\added{Given the exponential scaling of $N_C$ with dimension, future work should also investigate adaptive clustering strategies that exploit low-dimensional manifold structure in high-dimensional datasets, as well as multi-scale approaches that use coarser clusters in low-density regions. Furthermore, rigorous convergence analysis establishing quantitative bounds on the finite-$\sigma$ and finite-$N_C$ errors would strengthen the theoretical foundation of KGMM.} This will open up new possibilities for data-driven reduced-order modeling in climate science, fluid dynamics, and other areas where accurate score function estimation is crucial for capturing the stochastic behavior and long-term statistics of complex dynamical systems.

\added{All code used to generate the results in this manuscript is publicly available in open-source repositories, which include scripts to reproduce all figures and numerical experiments \footnote{\url{https://github.com/ludogiorgi/ScoreEstimation}, \url{https://github.com/ludogiorgi/ClustGen}}.}

\appendix

\section{\added{Technical Details and Hyperparameters}}
\label{app:hyperparameters}

\added{This appendix provides comprehensive technical details for all numerical experiments reported in Section 3, including neural network architectures, training hyperparameters, KGMM parameters, dataset sizes, decorrelation times, and random seeds. These details are essential for reproducibility and are organized by system.}

\subsection{\added{General Neural Network and Training Details}}

\added{All neural networks were implemented using the Flux.jl machine learning library in Julia. The architecture consists of fully connected (dense) layers with the Swish activation function defined as $\varphi(x) = x \cdot \sigma(x)$ where $\sigma(x) = 1/(1+\exp(-x))$ is the sigmoid function. The output layer uses a linear activation (identity function). Training was performed using the Adam optimizer \cite{kingma2014adam} with default hyperparameters ($\beta_1 = 0.9$, $\beta_2 = 0.999$, $\epsilon = 10^{-8}$) unless otherwise stated. The loss function is the mean squared error (MSE) between the predicted and target noise vectors $\bm{z}$:}
\begin{equation}\added{
\mathcal{L} = \frac{1}{N_C} \sum_{k=1}^{N_C} \left\| \bm{q}_{\theta}(\bm{C}_k) - \bm{q}_k \right\|_2^2,
}\end{equation}
\added{where $\bm{C}_k$ are the cluster centroids and $\bm{q}_k = -\frac{1}{|\Omega_k|} \sum_{i \in \Omega_k} \bm{z}_i$ are the target score estimates at each cluster.}

\subsection{\added{System-Specific Parameters}}

\subsubsection{\added{One-Dimensional Double-Well Potential}}
\begin{itemize}\added{
\item \textbf{Dimension:} $d = 1$
\item \textbf{Neural network architecture:} Hidden layers: [100, 50]
\item \textbf{KGMM parameters:} $\sigma \in [0.01, 0.05, 0.1, 0.5]$, $N_C = 31$, $\alpha = 10^{-3}$
\item \textbf{Training:} Learning rate $\eta = 10^{-3}$, Batch size $B = 16$, Epochs $n_{\text{epochs}} = 1000$
\item \textbf{Integration time step:} $dt = 0.01$
\item \textbf{Decorrelation time:} $t_d = 1.30$
\item \textbf{Uncorrelated samples:} $N_{\text{eff}} = 10^5$
}\end{itemize}

\subsubsection{\added{Reduced Triad Model}}
\begin{itemize}\added{
\item \textbf{Dimension:} $d = 1$
\item \textbf{Neural network architecture:} Hidden layers: [100, 50]
\item \textbf{KGMM parameters:} $\sigma = 0.01$, $N_C = 346$, $\alpha = 10^{-3}$
\item \textbf{Training:} Learning rate $\eta = 10^{-3}$, Batch size $B = 16$, Epochs $n_{\text{epochs}} = 1000$
\item \textbf{Integration time step:} $dt = 0.01$
\item \textbf{Decorrelation time:} $t_d = 0.62$
\item \textbf{Uncorrelated samples:} $N_{\text{eff}} = 10^5$
}\end{itemize}

\subsubsection{\added{Two-Dimensional Asymmetric Potential}}
\begin{itemize}\added{
\item \textbf{Dimension:} $d = 2$
\item \textbf{Neural network architecture:} Hidden layers: [128, 64]
\item \textbf{KGMM parameters:} $\sigma = 0.05$, $N_C = 725$, $\alpha = 10^{-3}$
\item \textbf{Training:} Learning rate $\eta = 10^{-3}$, Batch size $B = 64$, Epochs $n_{\text{epochs}} = 100$
\item \textbf{Integration time step:} $dt = 0.05$
\item \textbf{Decorrelation time:} $t_d = 3.00$
\item \textbf{Uncorrelated samples:} $N_{\text{eff}} = 10^5$
}\end{itemize}

\subsubsection{\added{Stochastic Lorenz 63}}
\begin{itemize}\added{
\item \textbf{Dimension:} $d = 3$
\item \textbf{Neural network architecture:} Hidden layers: [128, 64]
\item \textbf{KGMM parameters:} $\sigma = 0.05$, $N_C=754$, $\alpha = 10^{-3}$
\item \textbf{Training:} Learning rate $\eta = 10^{-3}$, Batch size $B = 64$, Epochs $n_{\text{epochs}} = 100$
\item \textbf{Integration time step:} $dt = 0.01$
\item \textbf{Decorrelation time:} $t_d = 0.30$
\item \textbf{Uncorrelated samples:} $N_{\text{eff}} = 10^5$
}\end{itemize}

\subsubsection{\added{Stochastic Lorenz 96}}
\begin{itemize}\added{
\item \textbf{Dimension:} $d = 4$
\item \textbf{Neural network architecture:} Hidden layers: [128, 64]
\item \textbf{KGMM parameters:} $\sigma = 0.05$, $N_C = 3818$, $\alpha = 10^{-3}$
\item \textbf{Training:} Learning rate $\eta = 10^{-3}$, Batch size $B = 64$, Epochs $n_{\text{epochs}} = 100$
\item \textbf{Integration time step:} $dt = 0.005$
\item \textbf{Decorrelation time:} $t_d = 0.19$
\item \textbf{Uncorrelated samples:} $N_{\text{eff}} = 10^5$
}\end{itemize}

\subsubsection{\added{Kuramoto--Sivashinsky Equation}}
\begin{itemize}\added{
\item \textbf{Dimensions:} $d \in \{4, 8, 16\}$ spatial discretization with $n_{\text{grid}}=128$ Fourier modes
\item \textbf{Subsampling:} Stride values $n_{\text{stride}} \in \{32,16,8\}$ for $d \in \{4,8,16\}$, respectively
\item \textbf{Data augmentation:} Circular shifts applied to each snapshot to produce 8 uncorrelated realizations per snapshot (augmentation factor $=8$)
\item \textbf{Neural network architecture:} Hidden layers: [128, 64]
\item \textbf{KGMM parameters:} $\sigma = 0.1$. Cluster counts: $N_C = 74{,}047$ ($d=4$), $N_C = 747{,}507$ ($d=8$), $N_C = 1{,}297{,}386$ ($d=16$), $\alpha = 10^{-3}$
\item \textbf{Training:} Learning rate $\eta = 10^{-3}$, Batch size $B = 64$, Epochs: $250$ ($d=4$), $200$ ($d=8$), $250$ ($d=16$)
\item \textbf{Integration time step:} $dt = 0.01$
\item \textbf{Decorrelation time:} $t_d = 1.5584$ ($d=4$), $10.6755$ ($d=8$), $1.5584$ ($d=16$)
\item \textbf{Uncorrelated samples:} $N_{\text{eff}} = 8 \times 10^5$ (obtained via 8x augmentation)
}\end{itemize}



\subsection{\added{Summary Table}}

\begin{table*}[t]\added{
\centering
\caption{Summary of parameters for all systems. Different values of $\sigma$ have been used for the Double Well (1D) system (see Section 2.4). For the KS equation, $N_{\text{eff}}$ includes an 8x augmentation factor from circular shifts.}
\label{tab:hyperparameters}
\small
\begin{tabular}{lccccccc}
\toprule
System & $d$ & $N_{\text{eff}}$ & $dt$ & $t_d$ & $\sigma$ & $N_C$ & Epochs \\
\midrule
Double Well (1D) & 1 & $10^5$ & 0.01 & 1.30 & -- & 31 & 1000 \\
Reduced Triad & 1 & $10^5$ & 0.01 & 0.62 & 0.01 & 346 & 1000 \\
2D Potential & 2 & $10^5$ & 0.05 & 3.00 & 0.05 & 725 & 100 \\
Lorenz 63 & 3 & $10^5$ & 0.01 & 0.30 & 0.05 & 754 & 100 \\
Lorenz 96 & 4 & $10^5$ & 0.005 & 0.19 & 0.05 & 3818 & 100 \\
KS ($d\!=\!4$) & 4 & $8 \times 10^5$ & 0.01 & 1.5584 & 0.1 & 74{,}047 & 250 \\
KS ($d\!=\!8$) & 8 & $8 \times 10^5$ & 0.01 & 10.6755 & 0.1 & 747{,}507 & 200 \\
KS ($d\!=\!16$) & 16 & $8 \times 10^5$ & 0.01 & 1.5584 & 0.1 & 1{,}297{,}386 & 250 \\
\bottomrule
\end{tabular}
}\end{table*}

\bibliography{references}
\end{document}